\documentclass[twocolumn,showpacs,preprintnumbers,amsmath,amssymb,aps,prl]{revtex4-1}
\usepackage{graphicx}
\usepackage{dcolumn}
\usepackage{bm}
\usepackage[shortlabels]{enumitem}
\begin{document}
.
\preprint{LA-UR-15-26915}

\title{Contribution of transverse modes to the dynamics of density fluctuations.}
\author{Giulia De Lorenzi-Venneri}
\affiliation{Theoretical Division, Los Alamos National Laboratory, 
Los Alamos, New Mexico 87545}
\author{Renzo Vallauri}
\affiliation{Istituto Sistemi Complessi, Consiglio Nazionale delle Ricerche, 50019 Sesto Fiorentino, (Firenze), Italy }
\affiliation{Dipartimento di Fisica, Universit{\`a} degli Studi di Trento, 30123 Povo, (Trento), Italy}

\date{\today}
\begin{abstract}
The transverse and longitudinal current correlation functions are evaluated in liquid and amorphous sodium by computer simulation. The study of the corresponding spectra as a function of the wavevector $k$ allows the evaluation of a dispersion curve for their peak position. The results are compared with recent  experimental findings [PNAS {\bf107}, 21985 (2010)], obtained by a new analysis of $S(k,\omega)$ measured by inelastic X-ray scattering in liquid and polycrystalline sodium. A substantial agreement between experimental and computer simulation results is found. The analysis of the line widths supports the Vibration-Transit theory picture of the dynamics in liquids.
\end{abstract}

\maketitle

The use of modern X-ray sources and spectrometers has allowed in recent years the measurement of the spectrum of density fluctuations with surprising great accuracy at very small frequencies, a range  where neutron spectroscopy has no access.  In particular the  Brillouin frequency range could finally be explored. One could study in great detail  the departure from the hydrodynamic regime, where the density fluctuations are sampled over a distance much larger than the interatomic spacing. In particular liquid metals have received great attention at the very early stage of the new experimental technique: results have been thoroughly reviewed by Scopigno, Ruocco and Sette in an exhaustive paper \cite{Scopigno-2005}, where an extensive list of references to the subject can also be found.

Parallel to the experimental effort there has been the attempt to present a comprehensive theoretical interpretation. This has been achieved by resorting to the Mori-Zwanzig~\cite{Zwanzig-1961,Mori-1965} approach. The method relies on the projection operator formalism to arrive to  a second order differential equation for the relevant correlation function. This correlation function contains the first order memory function,  for  which parameters have to be found by fitting  experimental data. An interpretation in terms of underlying physical processes is  provided by recognizing the effects of short time (free particle) dynamics and  structural relaxation. It is useful to remind that in this context the adopted projection operator method turns out to be equivalent to the assumption that the hydrodynamics point of view can be extended at the microscopic level if transport coefficients (diffusion, viscosity etc.) are regarded as both space and time dependent~\cite{Balzoppi-book}. In particular the memory function approach never involves quantities like generalized viscosities (shear and bulk), since the experimental spectra concern only the longitudinal current, whose spectrum is directly related with the experimentally available dynamical structure factor $S(k,\omega)$. In the hydrodynamics framework  in fact longitudinal and transverse currents are orthogonal.

An alternative approach, known as Vibration-Transit theory (V-T), has been more recently proposed by Wallace~\cite{DW-book2}. A liquid is treated as a disordered crystal with the same theoretical techniques valid for such a structure. The theory not only  gives a good account of most of the thermodynamic properties of liquid metals already in its zero order version, but it also provides a tool for describing the dynamical effects \cite{Chisolm-2001}. In this framework the system is pictured as vibrating  in what are referred to as random valleys, plus instantaneous transits among neighboring valleys. In this theory the longitudinal and transverse currents do not play a significant role even if their expressions can be calculated and indeed reported in detail \cite{arXiv:0702051v3}.

Molecular dynamics (MD) calculations have played an important role in confirming the physical interpretation since the early stages of computer simulation techniques. In a seminal paper~\cite{Rahman-1974} Rahman showed that liquid rubidium supports propagating density fluctuations for wave number $k <1.2 \AA^{-1}$ and found good agreement with neutron scattering results~\cite{Copley-1974}. Similarly the direct evaluation of the spectrum of the density fluctuations by computer simulation confirmed the accuracy of the V-T theoretical results for liquid sodium \cite{DeLorenzi-2005}, when compared with experimental data~\cite{Scopigno-2005b}.
\begin{figure} [t]
\includegraphics [width=0.45\textwidth]{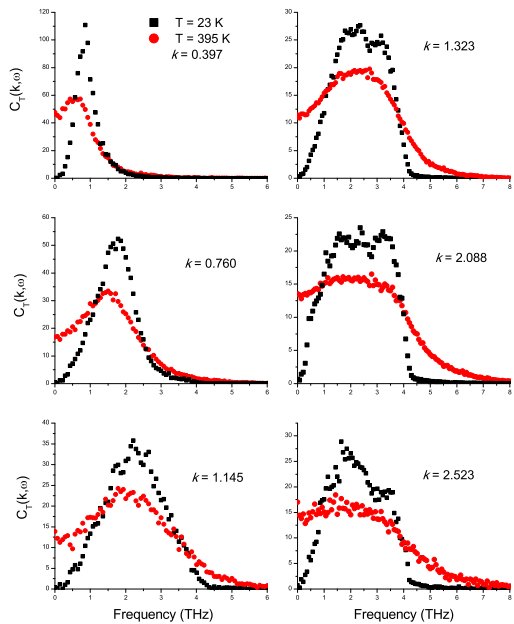}
\caption{Spectra of the transverse current at different wavevectors $k$, shown on each graph in $\AA^{-1}$ units. }
\label{fig1}
\end{figure}
\begin{figure} [t!]
\includegraphics [width=0.45\textwidth]{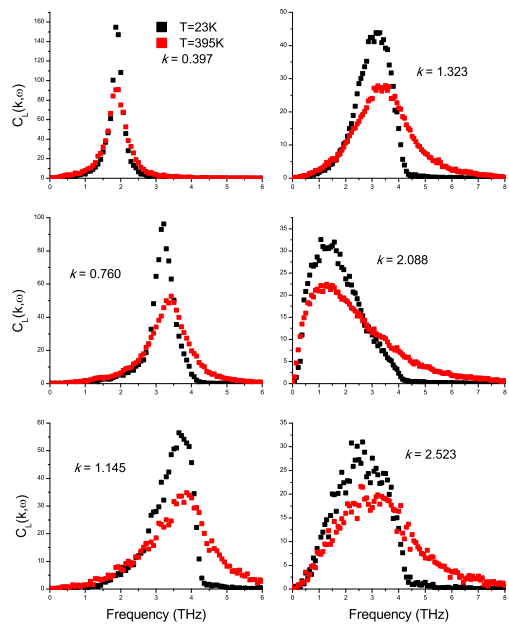}
\caption{Same as Fig.~\ref{fig1} but for the longitudinal current.}
\label{fig2}
\end{figure}

The fact that density fluctuations, as measured by neutron or e.m. scattering, reflects only the longitudinal current dynamics, became deep-rooted to such an extent that in dealing with the study of transverse current correlation by computer simulation, it was often reported that such a quantity cannot be measured by any experimental method. However the serendipity of molecular dynamics techniques opened new routes for the most comprehensive interpretation of the physical processes underlining the scattering events. In studying the dynamical properties of water, modeled by a new polarizable interaction potential, Balucani et al. \cite{Balucani-1996} pointed out that the spectrum of the longitudinal current shows the presence of two separate peaks at wavevector $k > 1.01$\AA$^{-1}$, one of them being coincident with the peak of the transverse current at similar wavevectors. Moreover, in the same range of wavevectors, the transverse current spectrum was found to display peaks at the same frequency of the longitudinal counterpart. As a consequence a wider representation of the inelastic X-ray scattering data of liquid water was suggested; a Damped Harmonic Oscillator (DHO) line shape, reflecting the transverse dynamics, was added to the viscoelastic model function which accounts for the central peak and the longitudinal dynamics \cite{Pontecorvo-2005}. A better agreement between experimental and fitted line shapes was claimed.

Quite recently Giordano and Monaco presented new experimental results of the spectrum of density fluctuations $S(k,\omega)$ of liquid sodium, obtained by inelastic X-ray scattering \cite{Giordano-2010}. The novelty and relevance of their paper arise from the fact that despite the firm belief that liquid metal dynamics was a well settled problem, completely understood in the generalized Langevin theoretical framework \cite{Scopigno-2005}, they analyze their experimental data in terms of a Lorentzian function for the central peak, \emph{plus two} DHO functions for the ``\emph{longitudinal and transverse excitations}".  To validate this assumption, they compared the experimental results for the liquid with similar findings, derived from the same  measurements performed on polycrystalline sodium. The correspondence of both longitudinal and transverse dispersion curves of liquid and polycrystalline phases reinforced the idea that liquid sodium can support transverse acoustic excitations beside the well-known longitudinal ones.

In the present paper we report computer simulation results of transverse and longitudinal current correlation functions and relative spectra. Our goal is to validate (refute) the following experimental outcomes : a) transverse modes, with frequency maxima close to the ones obtained by fitting the experimental $S(k,\omega)$, are present in liquid sodium; b) peak frequencies of these transverse modes change with the wavevector $k$ as found experimentally;  c) transverse modes become ill defined at increasing wavevectors.
The definitions of longitudinal and transverse currents and their corresponding correlation functions, $C_{L} (k,t)$ and $C_{T} (k,t)$, are well known \cite{Balzoppi-book} and there is no need to report them here. The Fourier transforms $C_{L}(k,\omega)$ and $C_{T}(k,\omega)$ of the normalized correlation functions will be the subject of our analysis.

MD simulations have been performed with 500 atoms in a cubic box of side $L$ with periodic boundary conditions. The well known Price, Singwi, Tosi \cite{Price-1970} potential model has been adopted along with the standard Verlet algorithm for the integration of the equation of motion with a time step $dt =0.25 \delta t$ where $\delta t =0.007$ps is the unit of time of our system. Two independent runs of 400.000 steps have been carried out; vectors  has been chosen of the form $ \bm{k}=(n,m,l)/L$, performing an average over all the possible triplets $(n,m,l)$ which realize the same vector amplitude $k$. The simulations have been done at two temperatures, namely $T = 395$K (just above melting) and $T =23$K, where the system has been brought after a rapid quenching in order to prevent crystallization. The comparison of the results at these different temperatures will give hints on the involved dynamical processes. A straight Fourier transform of the longitudinal and transverse current correlation functions for each  $k$ has been applied to obtain the requested spectra. These have been examined by searching the best Lorentzian lines fitting the data. The choice of the line shape has no particular physical reason in this context, but it is a simple tool to trace the maximum of the spectrum and its width. We have done the same analysis with a Gaussian line shape and no significance differences have been observed.

\begin{figure} [h!]
\includegraphics [width=0.40\textwidth]{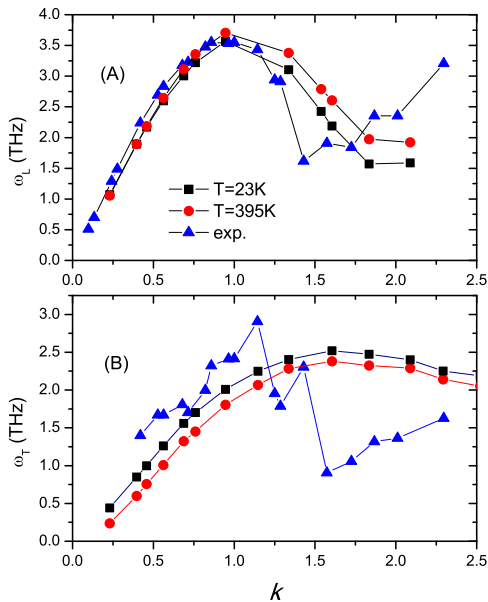}
\caption{Dispersion curve of the longitudinal (A) and transverse (B) current spectra. Triangles are the experimental data reported in \cite{Giordano-2010}. The wavevector $k$ is expressed in $\AA^{-1}$. Lines are guide for the eyes.}
\label{fig3}
\end{figure}

Typical spectra at the two temperatures and various wavevectors are reported in Fig.~\ref{fig1} and Fig.~\ref{fig2} for the transverse and longitudinal currents, respectively. A few comments are in order to shed light on the different behavior at the examined temperatures. 
First: the different value of the transverse spectra at $\omega = 0$ (very close to zero at $T=23$K, finite at $T=395$K). A simple explanation is provided by recognizing that as $k \rightarrow 0$ the normalized transverse current correlation function reads:
\begin{equation} \label{eq1}
\frac{C_{T}(k,t)}{C_{T}(k,t=0)} = \exp {(-\eta \frac{k^{2}}{\rho m} t)}.
\end{equation}
Where $\eta$ represent the shear viscosity. As a consequence,
\begin{equation} \label{eq2}
   C_{T}(k,\omega=0) = \int_0^\infty \! \exp{(-\eta \frac{k^{2}}{\rho m} t)} \, \mathrm d t = \frac {\rho m}{k^{2}} \frac {1}{\eta}~,
\end{equation}
 turns out to be very small at low temperature where the shear viscosity becomes extremely large. The same arguments apply at finite $k$ where, in a viscoelastic approximation, the shear viscosity $\eta$ is replaced by a wavevector dependent viscosity $\eta(k)$. A more detailed analysis can be found in \cite{Balzoppi-book} and {\cite{Balucani-1987}. 
Second: the appearance, at low temperature, of a second peak at larger wavevectors. As is evident from Fig.~\ref{fig1}, when $k > 1.13\AA^{-1}$ the transverse spectra show two distinct peaks which are very evident at $T=23$K. It has in fact been shown that, to a first approximation, transverse and longitudinal currents coincide with the velocity auto-correlation function (see Eq. 45 of \cite{arXiv:0702051v3}), whose spectrum at low temperature starts from zero due to the absence of diffusion processes and presents two well clear peaks. This feature is much less pronounced in the longitudinal current spectra since it is proportional to $\omega^{2}$, a fact that suppresses all small frequency characteristics.
Third: broadening of the peaks as temperature increases. This effect is due to the presence of more and more transits in the system at higher temperature. Such an effect has already been shown in the study of  density fluctuations and the corresponding spectra $S(k,\omega)$ \cite{DeLorenzi-2005}.
The peak position of the longitudinal and transverse current spectra (dispersion relation) are reported in Fig.~\ref{fig3}(A) and Fig.~\ref{fig3}(B), for the two examined temperatures along with the experimental findings provided in \cite{Giordano-2010}. The transverse dispersion relation has been obtained by fitting the spectrum by a single Lorentzian line at both temperatures.
\begin{figure} [h!]
\includegraphics [width=0.40\textwidth]{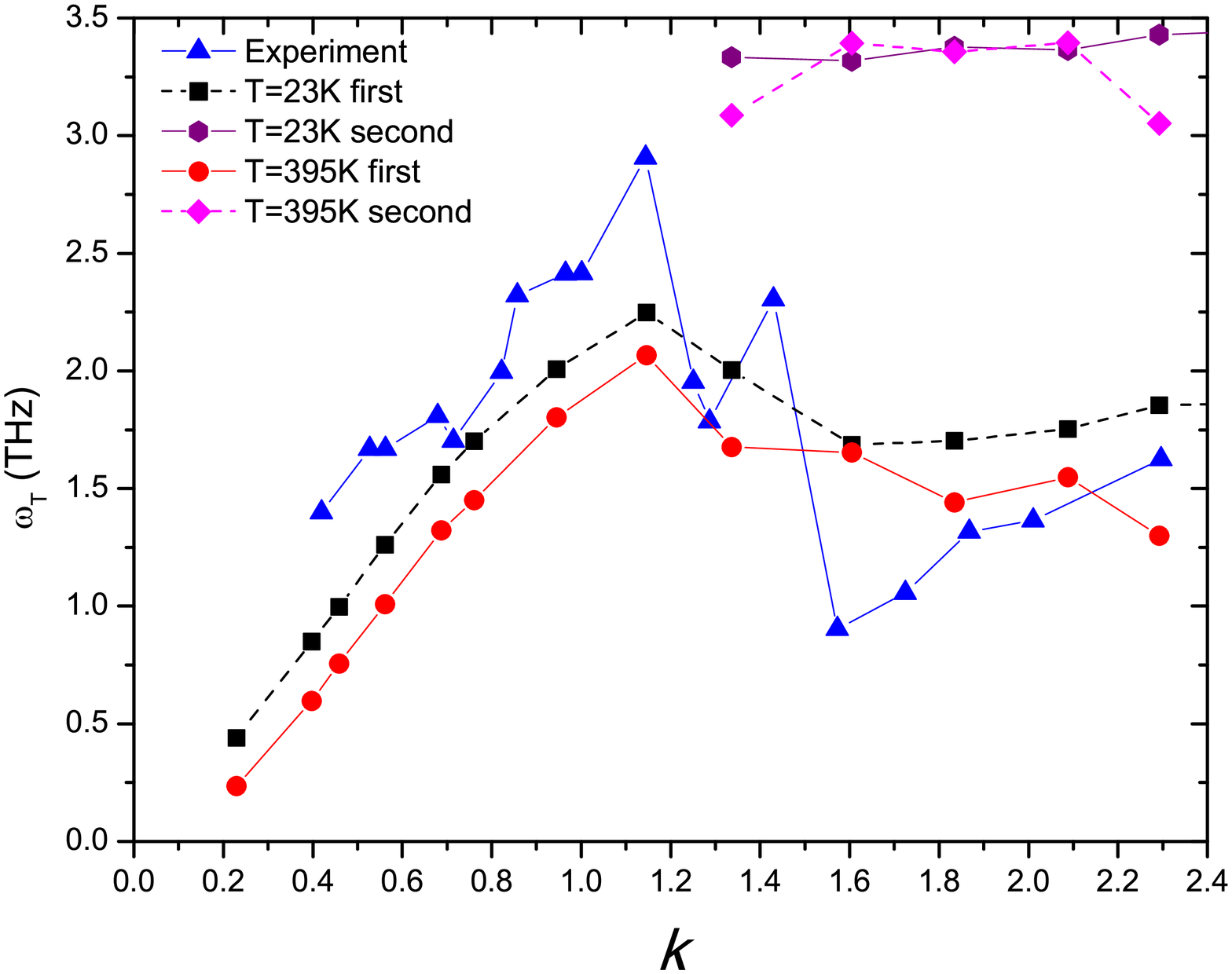}
\caption{Same as Fig.~\ref{fig3}(B), but fitting done with two Lorentzian lines for $k > 1.13\AA^{-1}$.}
\label{fig4}
\end{figure}

A perfect agreement between experimental and computer simulation  is evident for the longitudinal dispersion curves. The transverse dispersion curves are more complicated: for $0.2\AA^{-1} < k < 1.2\AA^{-1}$ the experimental transverse excitations are in reasonable agreement with the corresponding  simulation spectra. For $k > 1.2\AA^{-1}$ a substantial discrepancy is evident. As anticipated, in order to clarify this behavior we have performed the fitting of the spectra by two Lorentzian lines both at low and at high temperature. The results are shown in Fig.~\ref{fig4}. It appears that the experimental results keep track of the appearance of two transverse components at higher wavevectors.

\begin{figure} [h!]
\includegraphics [width=0.40\textwidth]{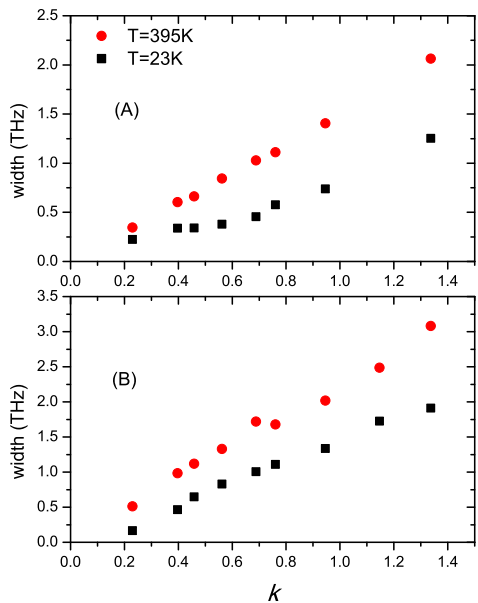}
\caption{Wavevector dependence of the current spectra width at the two temperatures: (A)
Longitudinal, (B) Transverse.}
\label{fig5}
\end{figure}

Finally it is interesting to observe the wavevector dependence of the width of the longitudinal and transverse current spectra, when both the spectra are fitted by a single Lorentzian line. Results are shown in Fig.~\ref{fig5}(A) and \ref{fig5}(B). There is a substantial increase of the width when the system is brought to the melting  temperature. As already discussed in \cite{DeLorenzi-2005} this result reflects the fact that at higher temperature transits occur in larger and larger numbers and destroy the coherence of the longitudinal and transverse oscillations.
Also, at both temperatures a linear $k$ dependence of the width is apparent. This is an indication that both systems are out of the hydrodynamics regime,  in which the  width would follow a $k^{2}$ dependence. 

In conclusion, MD calculations confirm the presence of transverse excitations in liquid sodium for wavevectors $ k > 0.2\AA^{-1}$. Their frequencies are in good agreement with those derived from the analysis of the experimental $ S(k,\omega)$ reported in~\cite{Giordano-2010}. The presence of two peaks in the transverse current spectra, as revealed by computer simulation, explain the experimental findings beyond $1.3\AA^{-1}$. This result is supported by the presence of a double peak  at low temperature. The large increase of the line widths of the longitudinal and transverse excitations at the higher temperature , as derived from MD simulations, gives  support to the description of the dynamical properties in simple liquids in terms of vibrations inside random valleys  and transits between them.

\acknowledgments{We thank G. Monaco for providing the experimental data and  P{\'a}l Jedlovszky for useful comments and suggestions.}
\bibliography{Currents}

\end{document}